\documentstyle[epsfig]{aipproc}

\begin{document}
\title{Spin dynamics in high-$T_C$ superconductors}

\author{Ph. Bourges$^1$, Y. Sidis$^1$, H.F. Fong$^2$, 
B. Keimer$^{2,3}$, L.P. Regnault$^4$,  
J. Bossy$^{5}$, A.S. Ivanov$^6$,
D.L. Milius$^{7}$,  and I.A. Aksay$^{7}$,}
\address{$^1$  Laboratoire L\'eon Brillouin, CEA-CNRS, CE Saclay, 
91191 Gif sur Yvette, France}
\address{$^2$ Department of Physics, Princeton University, Princeton, NJ 
08544, USA}
\address{$^3$ Max-Planck-Institut f\"ur Festk\"orperforschung, 70569 
Stuttgart, Germany}
\address{$^4$ CEA Grenoble, D\'epartement de Recherche Fondamentale sur 
la mati\`ere Condens\'ee, 38054 Grenoble cedex 9, France}
\address{$^5$ CNRS-CRTBT, BP 156, 38042 Grenoble Cedex 9, France }
\address{$^6$ Institut Laue-Langevin, 156X,  38042 Grenoble Cedex 9, 
France}
\address{$^7$ Department of Chemical Engineering, Princeton University, 
Princeton, NJ 08544 USA}
\maketitle

\begin{abstract}
Key features of antiferromagnetic dynamical correlations in high-$T_C$ 
superconductors cuprates are discussed. In underdoped regime, the sharp 
resonance peak, occuring exclusively in the SC state, is accompanied by 
a broader contribution located around $\sim$ 30 meV which remains above 
$T_C$. Their interplay may induce incommensurate structure 
in the superconducting state. 

\end{abstract}

\section*{Introduction}
Over the last decade, a great deal of effort has been devoted to show 
the importance of antiferromagnetic (AF) dynamical correlations for the 
physical properties of high-$T_C$ cuprates and consequently for   
the microscopic mechanism responsible for superconductivity\cite{model,pines}.
To elucidate  how these electronic correlations are relevant, it is then 
necessary to put the spectral weight of AF fluctuations on a more 
quantitative scale. Inelastic neutron scattering (INS) provides essential
information on this matter as it directly measures the full energy and 
momentum dependences of the spin-spin 
correlation function. Recently, efforts have been made to determine 
them in absolute units  by comparison with phonon scattering.
The following definition, corresponding to $1\over 3$ of the 
total spin susceptibility, is used\cite{recentprb}, 

\begin{equation}
\chi^{\alpha\beta} (Q,\omega)= -(g \mu_B)^2 {i\over{\hbar}} 
 \int_0^{\infty} dt \exp^{-i\omega t} <[S^{\alpha}_Q(t),S^{\beta}_{-Q}]>
\end{equation}

Our results are then directly comparable with both Nuclear Magnetic 
Resonance (NMR) results and theoretical calculations. 
Here, some aspects of the spin dynamics obtained in bilayer system 
will be presented in relation with recent  results reported 
by other groups\cite{incdai,mookinc}. However, it is before useful 
to recall the main features of magnetic correlations in the ${\rm YBa_{2}Cu_{3}O_{6+x}}$ (YBCO) system over doping 
and temperature\cite{rossat91,lpr,sympo,tony1,tony2,revue}.

\section*{Energy-dependences } 

\begin{figure} 
\centerline{\epsfig{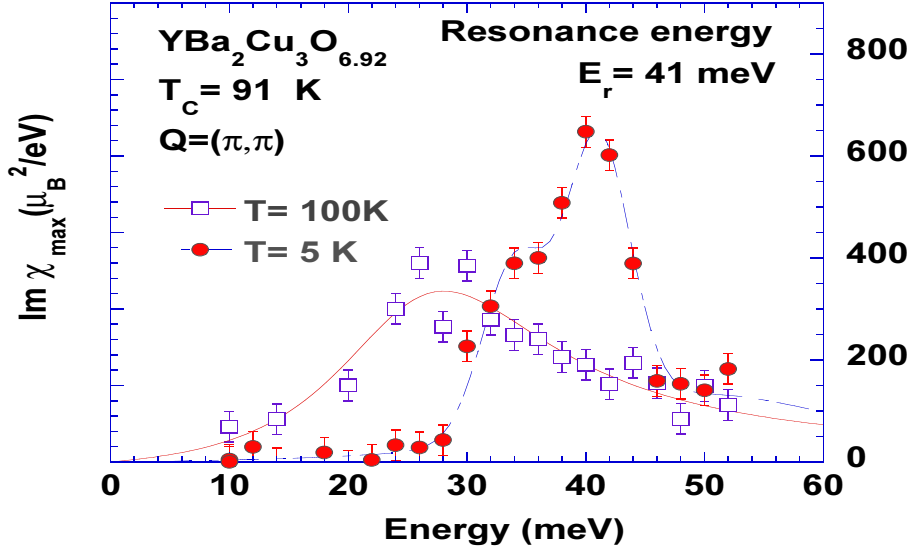} }
\vspace{10pt}
\caption{Low temperature (closed circles) and T=100 K (open squares) 
spin susceptibility at $Q=(\pi,\pi)$ in absolute units 
in YBCO$_{6.92}$ (2T-Saclay) (from [8]). }
\label{sqw692}
\end{figure}

 We first emphasize the 
energy dependence of the spin susceptibility at the AF wave vector,
 $Q_{AF}=(\pi,\pi)$, for x $\ge$  0.6 
(or $T_C \ge$ 60 K). 
$Im \chi$ in the normal state is basically 
well described in the underdoped regime by a broad peak centered around 
$ \simeq $ 30 meV (see Fig. \ref{sqw692})\cite{revue}. 
Upon heating, the AF spin
susceptibility spectral weight is reduced without noticeable 
renormalization in energy. Going into the superconducting state, 
a more complex line shape is observed essentially because a strong 
enhancement of the peak susceptibility occurs at some energy. 
This new feature is referred to as  the resonance peak, as it
is basically resolution-limited in energy (see e.g. 
\cite{rossat91,tony1,dai}). With increasing doping, 
the resonant peak becomes the major part of 
the spectrum\cite{revue}. At each doping, the  peak intensity at the
resonance energy is characterized by a striking temperature dependence 
displaying a pronounced kink at $T_C$ \cite{tony97,epl,dai,tonynew}.
Therefore, this mode is a novel signature of the unconventional 
superconducting state of cuprates which has spawned 
a considerable theoretical activity.
Most likely, the magnetic resonance peak is due to electron-hole pair 
excitation across the superconducting energy gap \cite{tony1,revue}.

The resonance peak may or may not be located at the same energy as the 
normal state peak. Fig. \ref{sqw692} displays a case where both occurs 
at different energies. However, at lower doping, these two features are 
located around similar energies, $\hbar\omega\sim$ 30-35 meV 
 for x $\sim$ 0.6-0.8\cite{revue,epl,tonynew}. Indeed,  the resonance 
energy more or less scales with the superconducting temperature 
transition\cite{revue,tony97,epl} whereas the normal state maximum does 
not shift much over the phase diagram for x $\ge$  0.6\cite{revue}.

Apart from the sharp resonance peak, the broad contribution 
(around $\sim$ 30 meV) is still discernible below $T_C$ as a shoulder,
 shown around $\hbar\omega \simeq$ 35 meV in Fig. \ref{sqw692}\cite{revue}.
In the superconducting state, the situation looks more complex as the 
low energy spin excitations are removed below a threshold, so-called  
spin gap\cite{rossat91,lpr,revue}, likely related to the superconducting 
gap itself. The non-resonant contribution has not received much 
attention so far. However, its spectral weight in the normal state
is important and may be crucial for a mechanism for 
the high-$T_C$ superconductivity based on antiferromagnetism\cite{pines}.

With increasing doping, the latter peak is continuously reduced: it 
becomes too weak to be measured in INS experiments in the 
overdoped regime YBCO$_7$\cite{lpr,tony1,tony2,revue}.
Using the same experimental setup and the same sample\cite{lpr,revue}, 
no antiferromagnetic fluctuations are discernible in the normal state 
above the nuclear background. Consistently, in the SC state, an isolated excitation around 40 meV is observed corresponding to the resonance peak. 
Above $T_C$, an upper limit 
for the spectral weight can be given\cite{tony2} which is about 4 times 
smaller than in YBCO$_{6.92}$\cite{revue}. Assuming the same momentum 
dependence as YBCO$_{6.92}$, it would give a maximum of the spin 
susceptibility less than 80 $\mu_B^2 / eV$ at $(\pi,\pi)$ in
 our units. Therefore, even though YBCO$_7$ may be near a Fermi 
liquid picture\cite{revue} with weak magnetic correlations, the 
spin susceptibility at $Q=(\pi,\pi)$ can still be $\sim$ 20 times 
larger than the uniform susceptibility measured by macroscopic 
susceptibility or deduced from NMR knight shift\cite{pines}. 

Therefore, $Im \chi$ is then naturally characterized in the superconducting 
state by two contributions having opposite doping dependences, 
the resonance peak becoming the major part of the spectrum with 
increasing doping. The discussion of Im$\chi$\ in terms of two 
contributions has not been emphasized by all groups\cite{dai}. 
However, we would like to point out that this offers a comprehensive 
description consistent with all neutron data in YBCO published so far.
In particular, it provides an helpful description of the puzzling 
modification of the spin susceptibility induced by 
zinc substitution\cite{yvan,tonyzn} by noticing that, on the one hand, 
zinc reduces the resonant part of the spectrum and, on the other hand, 
it restores AF non-resonant correlations in the normal state\cite{revue}.
Interestingly, the incommensurate peaks recently observed below the resonance 
peak in YBCO$_{6.6}$\cite{incdai,mookinc,arai} support the existence of two 
distinct contributions as the low energy incommensurate  
excitations cannot belong  to the same excitation as the 
commensurate sharp resonance peak.
Finally, these two contributions do not have  to be 
considered as independent and superimposed excitations: the occurrence 
of the sharp resonance peak clearly affects the full energy shape of 
the spin susceptibility\cite{rossat91,revue,tony97,epl,dai}.
We shall see below that the spin susceptibility q-shape is also modified 
below $T_C$.

\section*{Momentum-dependences}

In momentum-space, although both contributions are most generally peaked  
around the commensurate  in-plane wavevector $(\pi,\pi)$, they 
exhibit different q-widths. The resonance peak is systematically related 
to a doping independent q-width, 
$\Delta q^{reso} = 0.11 \pm 0.02$ \AA$^{-1}$\cite{sympo}  (HWHM), and 
hence to a larger real space distance, $\xi = 1/\Delta q^{reso} 
\simeq 9$ \AA\ in real space.
Recent data \cite{dai,epl,tonynew,arai} agree with that conclusion. 
In contrast, the non-resonant contribution exhibits a larger and 
doping dependent q-width, so that,
the momentum  width displays a minimum versus energy at  
the resonance peak energy\cite{sympo,epl,arai}. 

\begin{figure} 
\parbox{7.5 cm} {
\vspace*{-0.6 cm} 
\epsfig{file=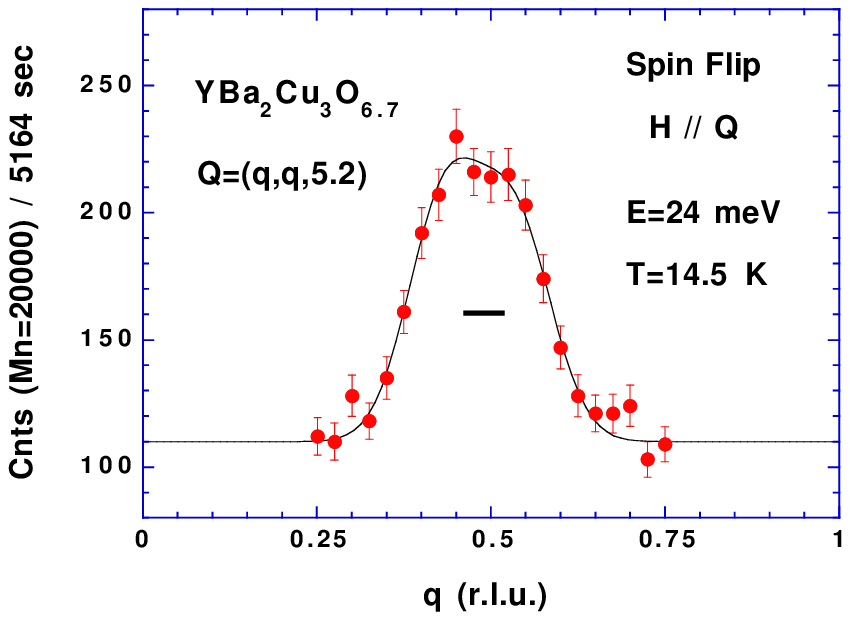,height=5 cm,width=7 cm} 
\vspace{10pt}
 }
 \parbox{7  cm} {
 \epsfig{file=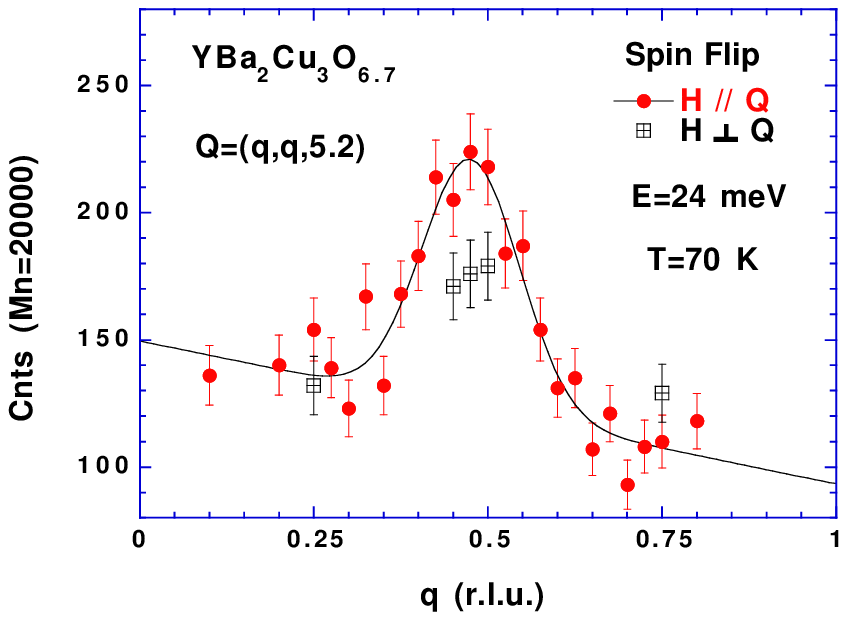,height=5 cm,width=7 cm} 
\vspace{10pt}
 }
\caption{Spin-flip Q-scans at 24 meV along the (11)) direction in 
YBCO$_{6.7}$ ($T_C$= 67 K): T= 14.5 K (right) and 70 K (left)  (IN20-Grenoble).
Polarized beam field was applied either parallel to Q (closed circles) or 
perpendicular to Q (dashed squares). The bar represents the q-resolution. }
\label{qscan}
\end{figure}

Recently, in the underdoped regime $x=0.6$, Dai et al\cite{incdai} 
reported low temperature 
q-scans at $\hbar\omega$= 24 meV which were  peaked at 
an incommensurate wavevector. Later, Mook et al\cite{mookinc} have 
detailed the precise position of these features, displaying a complex
structure of the incommensurability with a squared-like shape  
with more intense four corners at $Q=(\pi,\pi(1\pm\delta))$ and 
$Q=(\pi(1\pm\delta),\pi)$ with $\delta$= 0.21.
Interestingly, the energy where this structure is reported is 
systematically located in a small energy range below the resonance 
energy,  $E_r$= 34 meV\cite{dai,arai}. Further, this structure is only 
clearly observed at temperatures below $T_C$. In the normal state,
its existence remains questionable owing to background subtraction 
difficulties in such unpolarized neutron experiments\cite{mookinc,arai}.
A broad commensurate peak is unambiguously recovered  above 75 K in
polarized beam measurements\cite{dai}.
 
To clarify that situation, we have performed a polarized neutron 
triple-axis experiment on an underdoped sample YBCO$_{6.7}$ with 
$T_C$= 67 K\cite{tony97,tonynew}. The experiment on IN20 at the Institut
Laue Langevin (Grenoble) with a final wavevector $k_F$= 2.662 \AA$^{-1}$
(Experimental details will be reported elsewhere\cite{tonynew}). 
Fig. \ref{qscan} displays q-scans at $\hbar\omega$= 24 meV in the 
spin-flip channel at two  temperatures: T=14.5 K and T= $T_C$ + 3 K.
 The polarization analysis, and especially the comparison of 
the two guide field configurations (H // Q and H $\perp$ Q),
allows unambiguously to remove the phonon contributions\cite{tony2}.
Surprisingly, the magnetic intensity is basically found commensurate at both 
temperatures. Tilted goniometer scans have been performed to pass through 
the locus of the reported incommensurate peaks\cite{mookinc}: less 
magnetic intensity is measured there meaning that there is no clear 
sign of incommensurability in that sample. 
However, Fig. \ref{qscan} shows  different momentum shapes at both 
temperatures: a flatter top shape is found at low temperature indicating 
that the momentum dependence of the spin susceptibility evolves with 
temperature. Fig. 3 
underlines this point as it displays the temperature dependence of the 
intensity at both the commensurate wavevector 
and at the incommensurate positions (along the (310) direction as reported  
in Ref. \cite{incdai}). Two complementary behaviors are found: at 
the commensurate position, the peak intensity is reduced at $T_C$\cite{dai} 
whereas at the incommensurate position the intensity increases at a temperature 
which likely corresponds to $T_C$. As quoted by Dai {\it et al}\cite{incdai}, 
on cooling below $T_C$, the spectrum rearranges itself with a suppression 
at the commensurate point accompanied by an increase in intensity 
at incommensurate positions.

\begin{figure} 
\parbox{8.5 cm} {
\hspace*{-.5 cm} 
\epsfig{file=nodal.epsi,height=7 cm,width=8 cm} 
\vspace{10pt}
 }
 \parbox{6  cm} {
{\bf FIGURE 3.} Temperature dependence of unpolarized neutron intensity 
at $\hbar\omega$= 24 meV in YBCO$_{6.7}$
at commensurate position (dashed squares) and at the incommensurate 
wavevector (closed circles) ($k_F$= 3.85 \AA$^{-1}$, 2T-Saclay). 
Lines are guide to the eye 
to sketch for the
temperature evolution of the nuclear background. }
\label{inc-tdep}
\end{figure}

Therefore, even though our YBCO$_{6.7}$ sample does not exhibit well-defined 
incommensurate peaks, quite similar temperature dependences are
observed in both samples. Superconductivity likely triggers a redistribution 
of the magnetic response in momentum space, that may marginally result in an 
incommensurate structure in a narrow energy range. Interestingly,  
the sharp resonance peak simultaneously occurs. So that, superconductivity 
affects the spin susceptibility shape in {\bf both} the momentum and the 
energy spaces. Then, the interplay between the resonant and the 
non-resonant contributions may induce the incommensurate structure.  
In this respect, the magnetic incommensurability found 
in the La$_{2-x}$Sr$_x$CuO$_4$ system  would have a different origin 
as the wavevector dependence of Im$\chi$ in LSCO remains essentially the 
same across the superconducting transition\cite{mason}.  

\section*{concluding remarks}

Energy and momentum dependences of the antiferromagnetic fluctuations 
in high $T_C$ cuprates YBCO have been discussed. The sharp resonance peak
occurs exclusively below $T_C$. It is likely an intrinsic feature of 
the copper oxides as it has been, recently, discovered in 
$\rm Bi_2 Sr_2 Ca Cu_2 O_{8+\delta}$\cite{resobsco}. This 
resonance peak is accompanied in underdoped samples by 
a broader contribution remaining above $T_C$.

\end{document}